\documentclass[letterpaper,prb,twocolumn,english]{revtex4-1}
\usepackage[pdftex]{graphicx}
\usepackage{bm}
\usepackage{epstopdf}
\usepackage{amsmath}
\usepackage{color}
\newcommand{\be}{\begin{equation}}
\newcommand{\ee}{\end{equation}}
\newcommand{\bdm}{\begin{displaymath}}
\newcommand{\edm}{\end{displaymath}}
\newcommand{\bit}{\begin{itemize}}
\newcommand{\eit}{\end{itemize}}
\newcommand{\ben}{\begin{enumerate}}
\newcommand{\een}{\end{enumerate}}

\begin{document}

\title{Autonomous movement of chemically powered vesicle}
\author{\small  Shivam Gupta$^1$, Sreeja K. K.$^2$ and Snigdha Thakur$^1$}
\affiliation{ \small 1.Department of Physics, Indian Institute of Science Education and Research Bhopal, Bhopal, India\\ \small 2. Department of Physics, Indian Institute of Technology Madras, Chennai, India}

\begin{abstract}

\noindent A mechanism for self propulsion of deformable vesicle has been proposed, vesicle moves by sensing the self-generated chemical gradient. Like many molecular motors they suffer strong perturbations from the environment in which they move as a result of thermal fluctuations and do not rely on inertia for their propulsion. Motion of the vesicle is driven by an asymmetric distribution of reaction products. The propulsive velocity of the device is calculated as well as the scale of the velocity fluctuations. We present the simulation results on velocity of vesicle, reorientation of vesicle, and shape transformation of vesicles.      

\end{abstract}
\maketitle{}

\section{Introduction}   \label{intro}

Nondiffusive transport of objects such as vesicles, colloids and aggregates plays a fundamental role in biological systems. What is most important for these small length scale objects for motility is some degree of asymmetry. Several mechanisms that can lead to such motion with the help of asymmetry have been predicted~\cite{golestanian:05,nardi:99,sumino:05,kay:07,anderson:83}.   One important mechanism responsible for the propulsion is the conversion of chemical energy into directed motion.  At cellular level, nano-sized biological molecules  are transported with remarkable efficiency by biological motors such as kinesin through chemically induced symmetry-breaking conformational changes.~\cite{vale:00,yildiz:06} In this paper we report the autonomous propulsion of  vesicle using the above, as  an underlying mechanism.

Inspired by nature synthetic structures  have been designed that uses chemical, light or other form of energy to perform directed motion.~\cite{kay:07} Two different strategy are followed for the construction of such motors. In one biology is copied and artificial molecular machines are  constructed such that they use chemical energy to drive the conformational changes and hence perform directed motion~\cite{shin:04,shapere:87}. For the other route the propulsion rely on asymmetric chemical reactivity instead of asymmetric conformational dynamics. The most widely studied example of this type of motors are bimetallic nanorods~\cite{sen:04,wang:13}, Janus particles~\cite{golestanian-1:07}, sphere dimer motors~\cite{ozin:10,ruckner:07,snigdha:12}, polymers~\cite{wu:13} The above cited example of physically asymmetric self-proplled objects assume that the particle shape is unchanged during the motion. However, in reality many self-propelled objects may change their shape depending on the velocity, environment or interaction with other objects.  Recently there have been  reports of autonomous propulsion for deformable objects with propulsive effect created by the physical asymmetry.~\cite{wilson-12,howse-12,ohta:09,sano-13,miura:09,giardini:03} Ohta and coworkers have theoretically investigated the individual and collective motion of deformable objects exhibiting self-propulsion.~\cite{ohta:09,ohta-12} They have found a bifurcation from a straight line to a circular motion in case of an isolated particle. In a different work  Giardini and coworkers have shown the propulsion of large artificial lipid vesicles due to the forces generated by actin in a motile comet tail.~\cite{giardini:03} Another very recent work by Wilson and group have experimentally demonstrated the autonomous movement of platinum-loaded stomatocytes.~\cite{wilson-12} These polymeric analogue of liposomes are created by the diblock copolymers and exhibit remarkably high average directed velocity even in presence of small amount of fuel- which is why they can be seen as ideal drug delivery devices. 

Here, we illustrate the self-propulsion of a chemo-mechanically active vesicle. A molecular description
of chemically powered directed motion was obtained in the triangulated vesicle model with localized asymmetric catalytic reactions. Vesicle consists of linked catalytic (C) and non-catalytic (N) vertices, which then is immersed in a solvent containing reactive A species. An irreversible reaction $A+C \rightarrow B+C$ takes place at all the catalytic vertex, which in turn gives rise to nonequilibrium gradient of B species. The combination of nonequilibrium gradients and potential asymmetry gives rise to the directed motion, similar to nanodimers.~\cite{ruckner:07} Here we discuss the factors that control the vesicle velocity and show how vesicles can be designed to have velocities that are larger than the thermal fluctuations, leading to easily observed directed motion. Such directed motion is essential for the applications where such propelled vesicles may possible be used as a carrier of target drug delivery or to perform specific task.

The paper is organized as follows. In Section~\ref{sec-model} we describe the particle-based mesoscopic model for the vesicle motors and solvent. Section~\ref{sec-prop} presents various  vesicles shape obtained in our model and their propulsion behavior. The efficiency of the vesicle motors of different shapes has been discussed in Section~\ref{sec-eff}. Finally, the conclusions of our study are presented in Section~\ref{sec-conc}.

\section{Simulation model}
\label{sec-model}
Self-propulsion of the fluid vesicle is simulated using a  mesoscopic approach, which combines a particle based hydrodynamics model for solvent~\cite{kapral_adv_chem_phys_2008,gompper:09,malevanets-99,malevanets-00} and a mesoscale~\cite{ho-89}, dynamically triangulated surface model for the membrane~\cite{nelson-book}. Similar model has been used previously to study the dynamics of fluid vesicle in capillary and shear flows.~\cite{noguchi-05,gompper-05}

\subsection*{Triangulated-surface model for the vesicle}

The vesicular membrane is modelled using dynamics triangulation method in which the continuum surface is replaced by a descretized surface defined by vertices, triangles and links. A spherical topology with the triangulation thus consists of $N_v$ vertices connected by $N_L=3(N_v-2)$ links,  $N_T= 2(N_v-2)$ triangles. Such  description of the membrane is extremely useful when only the gross features of the structure and interactions are important. The curvature elastic energy of the vesicle responsible for the shape and fluctuations is given by Canham Helfrich Model~\cite{helfrich-73}:
\be
H_c = \frac{\kappa}{2}\int(c_1+c_2)^2da
\ee
where, $c_1$ and $c_2$ denote the two principle curvatures and $\kappa$ is the bending stiffness of the membrane.
For the triangulated surface we use a discretized form of this Hel-
frich Hamiltonian proposed by Itzykson~\cite{nelson-book,gompper-05}.

Unlike the Monte Carlo simulations of triangulated membrane we use Stillinger Weber type potential~\cite{weber-85} for bonding ($U_b$) the vertices and also for excluded volume interaction ($U_{ex}$). The attractive bonding potential between the connected vertices is give by
\be
U_{b}(r)= 
\begin{cases}
    \frac{\gamma \exp[1/(r_{c0}-r)]}{r_{max}-r},& \text{if } r > r_{c0}\\
    $\vspace{0.1in} $
    0 & \text{if } r \leq r_{c0}
 \end{cases}
\ee
where $r = |\bf{r_{ij}}|$ is the bond length between vertices $i$ and $j$. The repulsive excluded volume interaction is incorporated by:
\be
    U_{ex}(r)= 
\begin{cases}
    \frac{\gamma\exp[1/(r-r_{c1})]}{r-r_{min}},& \text{if } r < r_{c1}\\
   $\vspace{0.1in} $
    0 & \text{if } r \geq r_{c1}.
\end{cases}
\ee
We use $r_{min}=0.67a_0$ and $r_{max}=1.33a_0$ as the minimum and maximum possible distance between vertices. The time evolution of the system is governed by hybrid molecular dynamics-multiparticle collision (MD-MPC) dynamics~\cite{kapral_adv_chem_phys_2008}. 

The triangulated membrane acquires its lateral fluidity from the bond flip mechanism, where the tethers can be flipped between the two possible diagonals of two adjacent triangles at time interval $\tau_{BF}$.~\cite{gompper-05} Flipping between catalytic and non-catalytic part of the vesicle is not allowed. 
\subsection*{Multiparticle collision dynamics of the solvent}

To bridge the large length and timescale gaps in the vesicle dynamics a mesoscopic model for the solvent is required.
In multiparticle collision dynamics, the solvent is described by $N_s$ pointlike particle of mass $m$, moving in the simulation box of size $L_x\times L_y\times L_z$. In this scheme, the evolution consists of series of streaming and multiparticle collision steps. In the streaming step, the dynamics is evolved by Newton's equations of motion governed by forces determined from the total potential energy $V({\bf r}^{N_v},{\bf r}^{N_s})$ of the system~\cite{snigdha-14}.  In the collision steps, which occur at time intervals $\tau$, the point-like solvent particles are sorted into cubic cells with cell size $a_0$. The choice of cell size is such that the mean free path $\lambda < a_0$. Multiparticle collisions among the solvent molecules are performed independently in each cell, which results in the post collision velocity of solvent particle $i$ in cell $\xi$ being given by
${\bf v_i^\prime} = {\bf V}_\xi + \hat\omega_\xi ({\bf v_i} - {\bf V_\xi})$, where ${\bf V_\xi}$ is the center-of-mass  velocity of particles in cell $\xi$ and $\hat \omega_\xi$ is a rotation matrix. In order to ensure Galilean invariance for systems with small $\lambda$, a random grid shift is applied in each direction of the simulation box~\cite{kroll-01,kroll-03}. The method described above conserves mass, momentum and energy and accounts for the hydrodynamic interactions and fluid flow fields~\cite{kapral_adv_chem_phys_2008,gompper:09}, which are important for the dynamics of the active fluid vesicle.

\subsection*{Membrane and solvent interaction}
\begin{figure}
\begin{center}
\includegraphics[width=0.6\columnwidth]{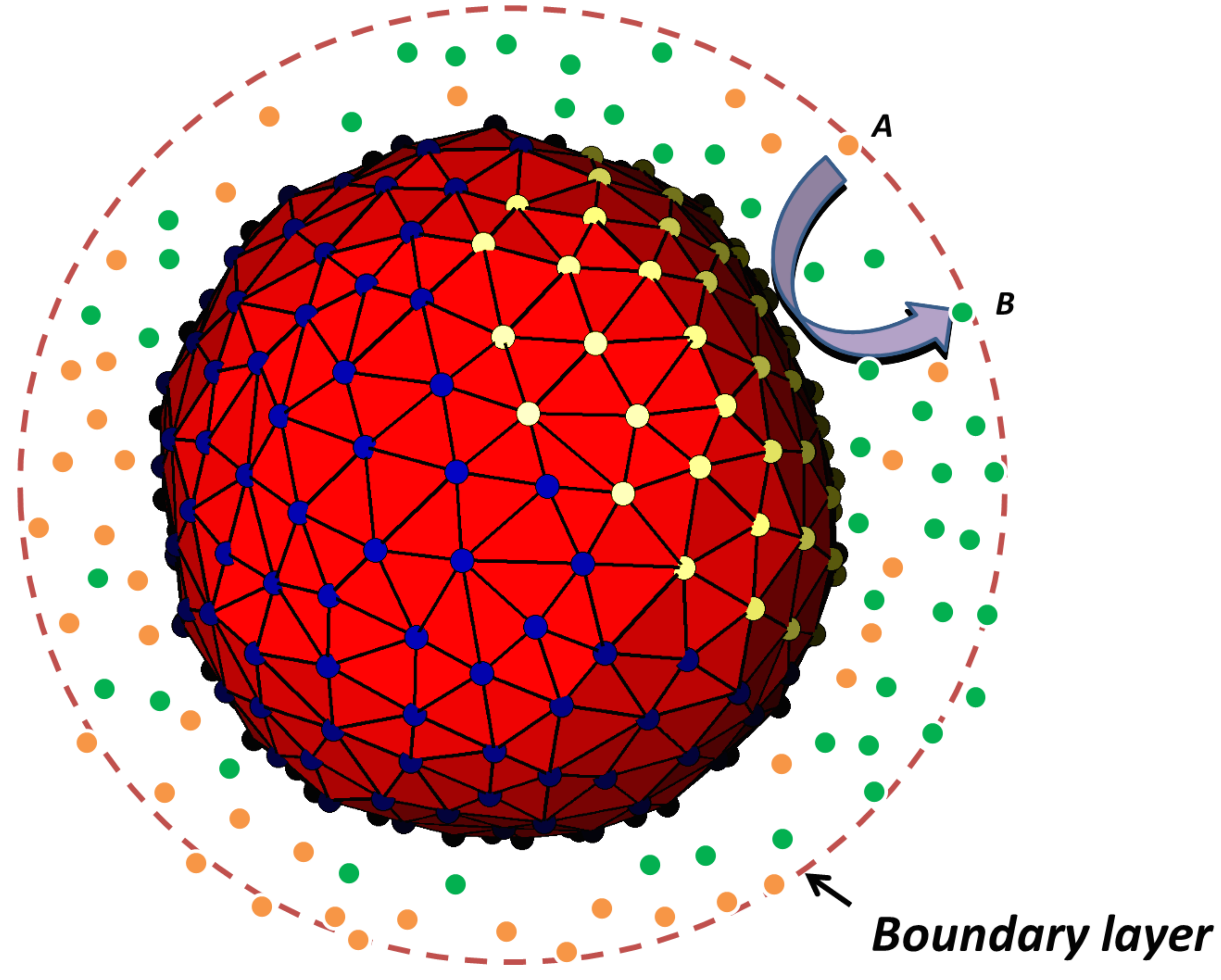}
\caption{\label{fig:reaction}A chemically active vesicle composed of catalytic (yellow)  and non-catalytic (blue) vertices. The figure depicts the chemical reaction that leads to conversion of fuel $A$ (orange) to product $B$ (green) molecule within the boundary layer, hence producing the inhomogeneous distribution of chemical species. The boundary layer around the vesicle
within which the intermolecular forces act is also shown.} 
\end{center}
\end{figure}

Theoretical descriptions of the vesicle motion arises from self-diffusiophoresis~\cite{ruckner:07,golestanian:05,golestanian:07}, which in this case is based on phoretic motion of the vesicle in external inhomogeneous chemical field. To model this, the vesicle is immersed in a fluid, where $N_s$ solvent particles in the MPC model is divided in to $A$ and $B$ kind of particles. A fraction of membrane vertices are considered to be chemically active where the chemical reaction $A \rightarrow B$ takes place,  producing an inhomogeneous concentration of solute molecules  on the scale of the characteristic size of the particle, which in turn is responsible for the propulsion of vesicle (See fig.~\ref{fig:reaction}). Such chemically active vertices are called catalytic (C) vertices and others are Non-catalytic (N) chemically inactive vertices. The $A$ and $B$ solvent molecular species interact with all the vesicle vertices through repulsive Lennard-Jones potentials,
\be
\label{eq:LJ}
V_{\beta S}(r) = 4\epsilon _\beta \left[ \left(\frac{\sigma}{r}\right)^{12} - \left(\frac{\sigma}{r}\right)^6 + \frac{1}{4}\right], r\leq r_c,
\ee
with the cutoff distance $r_c=2^{1/6}\sigma$. We use the notation $V_{\beta S}$, where $S=C,N$ and $\beta=A,B$
to denote various interactions between solvent and membrane. In particular, we take $V_{AC} = V_{BC} = V_{AN}$,
which are characterized by the energy and distance parameters $\epsilon_A$ and $\sigma$, respectively;
however, interactions between the $N$ vertex and $B$ molecules, $V_{BN}$, have energy parameter $\epsilon_B$. Hence the $B$ molecules produced in the reaction on catalytic sphere, react differently with catalytic and noncatalytic monomers and this asymmetry is an important element in the self-propulsion mechanism for the vesicle~\cite{ruckner:07,snigdha:10}. 
In order to mimic the fluxes of reactive species into and out of the system to drive it out of equilibrium and lead to establishment of a nonequilibrium stead state, $B$ molecules are converted back to $A$ molecules when they diffuse far enough away from the vesicle.

\subsection*{Simulation parameter}

Other simulation details are as follows: all quantities are reported in dimensionless LJ units based on energy $\epsilon_A$, mass $m$, and distance $\sigma$ parameters. The vesicle move in a solvent of $A$ and $B$ molecules within a cubic box of volume $V$ and linear dimension 30 with periodic boundary conditions. The MPC rotation angle was fixed at $\phi = 90^\circ$. The average number density of solvent outside the vesicle was kept at $\rho_{out} \approx 10$, whereas the inside solvent number density $\rho_{in}$ was varied from 1 to 10 to vary shape of the vesicle. The masses of both A and B species are $m = 1$ and the masses of the membrane vertices were adjusted according to their diameters, $d_v = 2\sigma$ to ensure the density matching between vertex and solvent. The system temperature was fixed at $T = 0.26$. Newton's equation of motion is integrated using velocity Verlet algorithm, with MD time step $\Delta t = 0.005$. The MPC time interval is $\tau = 0.25$ and the time interval for bond flip is $\tau_{BF}=0.1$. We have probed various sizes of the vesicle where $N_v$ varies from 110 to 677. Unless specified the value of $\kappa = 5$ and $\gamma = 80$. For the bond and excluded volume interaction, we use the cutoff lengths to be $r_{c0}=1.15$ and $r_{c1}=0.85$. The LJ parameters were chosen to be $\epsilon_A = 1.0$, while $\epsilon_B$ varies from 0.01 to 0.2.

\section{Membrane vesicle propulsion} 
\label{sec-prop}
\begin{figure}
\begin{center}
\includegraphics[width=1.0\columnwidth]{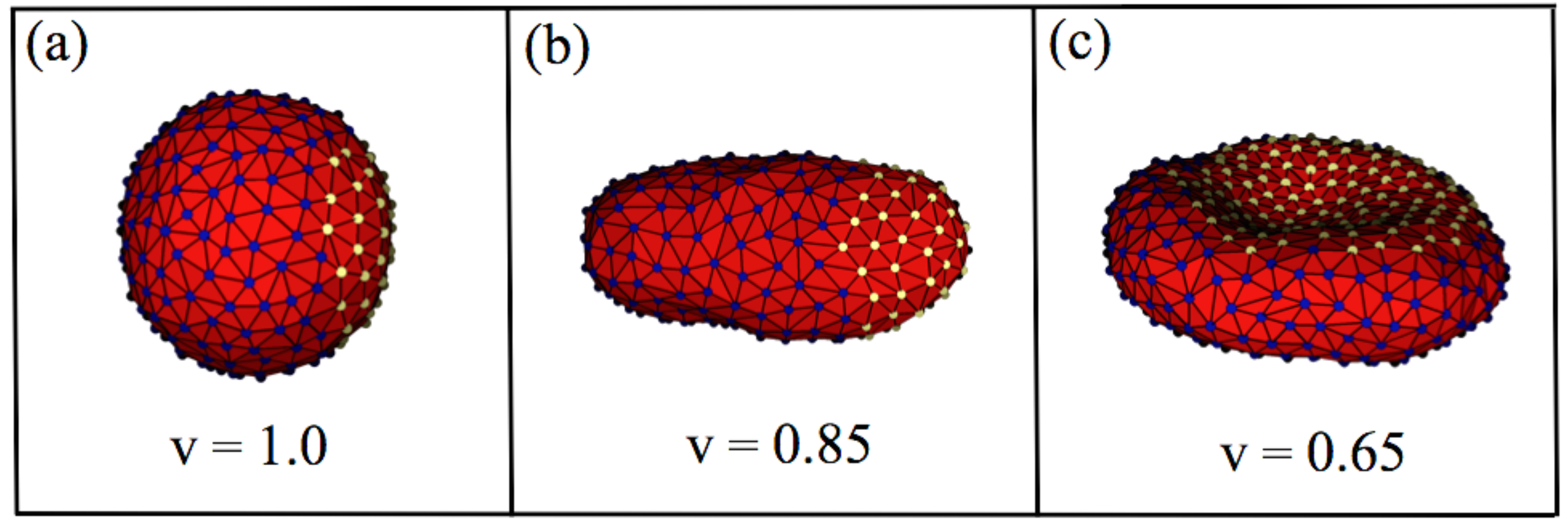}
\caption{\label{fig:shapes}Shape transformation of membrane vesicle as a function of the reduced volume $v$; (a) sphere (b) ellipsoid and (c) discoid} 
\end{center}
\end{figure}

We investigate the self-propulsion of vesicle of different shapes- spherical, ellipsoidal and discoidal vesicles. These various shapes are generated by changing the number of solvent particle density enclosed by the vesicle which in turn affect the area-to-volume ratio leading to a shape transformation.  
The shape characterization is done in terms of dimensionless variable, the reduced volume $v$, defined by $v= \frac{V_{ves}}{4\pi R_0^3/3}$, where $R_0$ is the radius of a sphere with the same surface area as that of the vesicle and $V_{ves}$ is the volume of the vesicle. For $v = 1$, the shape is necessarily spherical, which we obtained by keeping inside and outside solvent number density same. On reducing  inside solvent density we get the prolate/ellipsoidal shape at reduced volume $v \approx 0.85$, further reduction of reduced volume to $v \approx 0.65$ leads to discocytes vesicles. Fig.~\ref{fig:shapes} shows different shapes obtained in our simulations at different reduced volume.
 
\subsection*{Spherical vesicle}

\begin{figure}[h]
\begin{center}
\includegraphics[width=0.9\columnwidth]{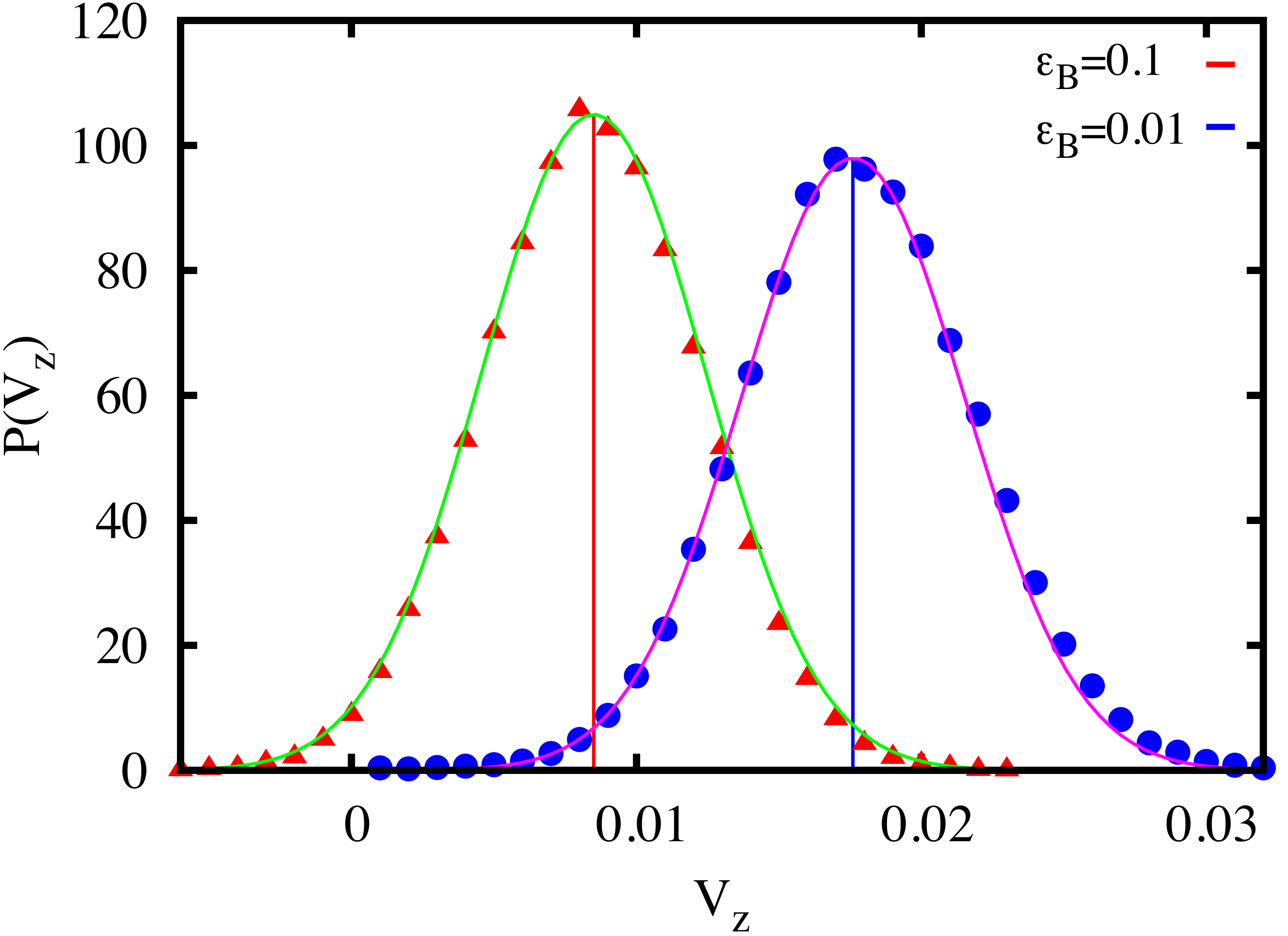}
\caption{\label{fig:lj_vel}Plot of $P(V_z)$, the probability density of $V_z$ for vesicle with $N_v=302$. Here, the energy parameter $\epsilon_A=1.0$ is used for both the curves. Green and pink solid lines are maxwell boltzmann fit to data. $\langle V_z \rangle = 0.0085$ for $\epsilon _B = 0.1$ and $\langle V_z \rangle = 0.0176$ for $\epsilon _B = 0.01$}.
\end{center}
\end{figure}

The self-propulsion of a membrane vesicle occurs as a result of the nonequilibrium concentration gradient produced by the chemical reaction at the catalytic vertices of the vesicle, in conjunction with the different intermolecular forces between the A and B chemical species and the vertices. The ability of the self-propelled vesicle to execute directed motion is determined by monitoring the mean value of the center of mass velocity of the vesicle motor, projected along the axis between the center of mass of the vesicle to the center of mass of catalytic vertices: $V_z =  \langle \langle {\bf V}(t) \cdot {\bf \hat R}(t) \rangle \rangle$, where the double angular bracket denote the average over time and realization. ${\bf \hat R}(t)$ is the unit vector along the above said axis.  An isolated vesicle motor undergoes self-propelled motion in the given direction with average velocity $\langle V_z \rangle$; however, such small motors are also subjected to strong thermal fluctuation that lead to a distribution of propagation velocities. This distribution has been shown to be closely approximated by a Boltzmann distribution~\cite{ruckner:07,snigdha:10,snigdha:12} with mean $\langle V_z \rangle$. The magnitude and nature of the directed motion is strongly influenced by the vesicle solvent interactions. Fig.~\ref{fig:lj_vel} shows the normalized probability distribution $P(V_z)$ for the system with $N_v=302$ with $C_f = 25\%$ ($C_f$ is percentage fraction of total vertex that are catalytic) for different values of energy parameter $\epsilon_B$.

As stated before asymmetry is a crucial component  for the self-propulsion for the vesicle, in the current system mechanism this  asymmetry was supplied by the existence of a strong product molecule $B$ concentration gradient and  difference in interaction potential of the product $B$ molecule with catalytic and non-catalytic vertex. In fig.~\ref{fig:lj_vel} we calculate the average directed velocity of the vesicle $\langle V_z \rangle$ for interaction potential strength $\epsilon_B$ equal to $0.01$ and $0.1$ keeping $\epsilon_A=1.0$. The figure clearly illustrate better propulsion for larger difference of the interaction strength i.e. for higher asymmetry.  

\begin{figure}
\begin{center}
\includegraphics[width=0.8\columnwidth]{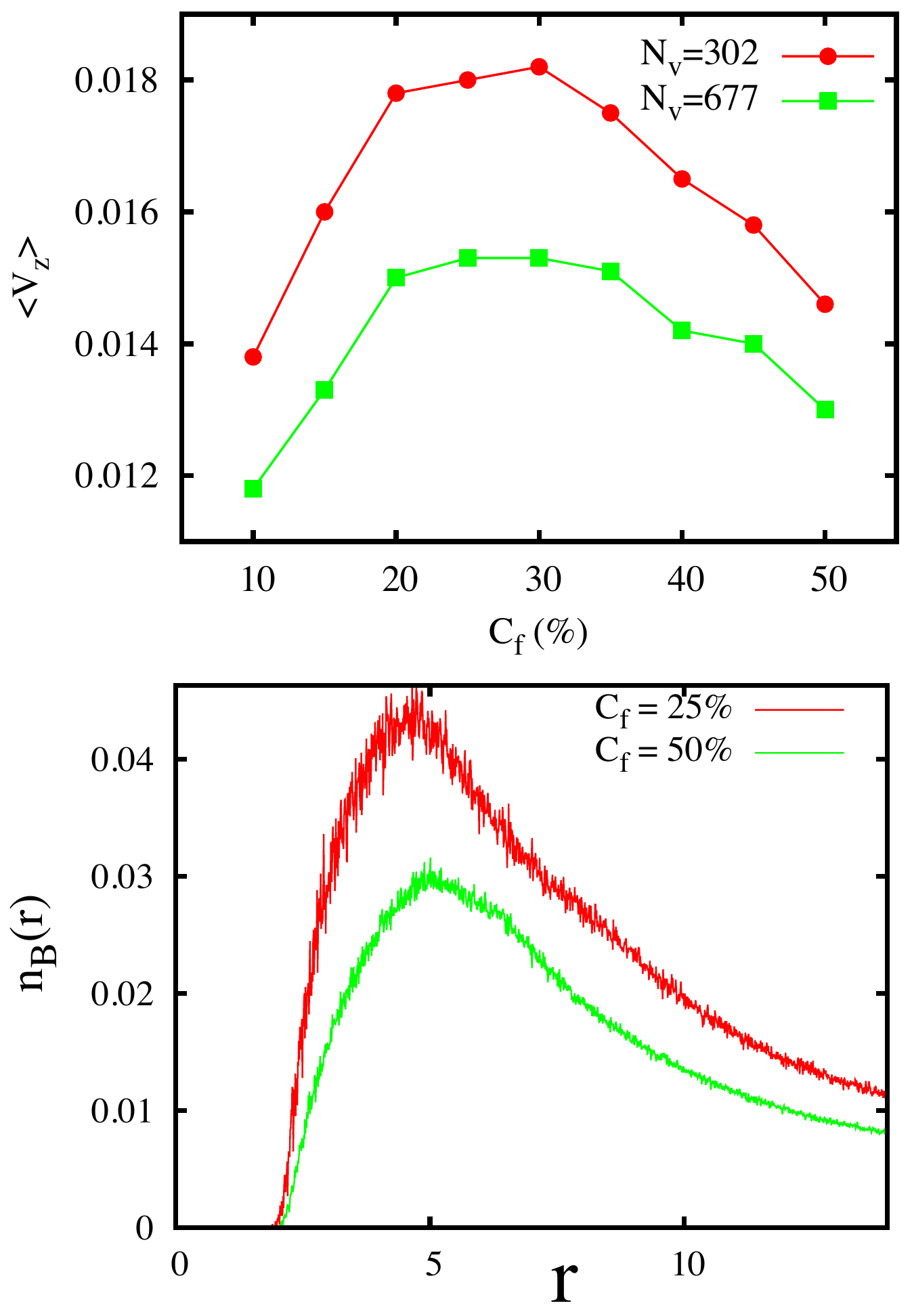}
\caption{\label{fig:cfrac} Top: Variation of average directed velocity of the vesicle $\langle V_z \rangle$ as a function of percentage of catalytic vertices $C_f$ for  $N_v=302$(red circles) and $N_v=677$ (green square). Bottom: The local steady state concentration of $B$ particles around the catalytic center of mass,$n_B(r)$, for  $C_f = 25\%$ and $50\%$.}
\end{center}
\end{figure}

Another important element for vesicle propulsion is the local steady state concentration of species $B$ in the vicinity of the catalytic $C$ vertices.  In order to have a vesicle with maximum possible directed movement it is important to have a optimum number of catalytic vertices on it. Having too many or too less catalytic vertex will not be able establish the sufficient nonequilibrium B concentration gradient around the vesicle; hence, directed vesicle propulsion will not be possible. Fig.~\ref{fig:cfrac} (Top) shows the average velocity of vesicle $\langle V_z \rangle$ as a function of percent fraction of catalytic vertices $C_f$. From the figure it is clear that $C_f \approx 25\% $ catalytic gives maximum directed velocity $\langle V_z \rangle \approx 0.018$ for $N_v = 302$, which are comparable to the previously reported sphere dimer motors~\cite{yuguo-08}. This is further supported by higher strength of the local steady state concentration gradient of species $B$ in the vicinity of the catalytic vertices shown in Fig.~\ref{fig:cfrac} (Bottom), for $C_f = 25\%$ and $50\%$. In figure $r$ is the distance of solvent $B$ molecules from the center of mass of catalytic vertices.

The nanoscale objects  will not simply move ballistically in a given direction due to strong fluctuations from surrounding medium~\cite{golestanian-1:07}. Instead, the motion will be ballistic at short times,  but at longer times, the motion will change to random walk, in which the spans of directed motion will be interrupted by random change in direction. The effective diffusion coefficient  can then be calculated by the velocity autocorrelation function as 
\be
D_e = \frac{1}{d} \int_0^{\infty}dt \langle {\bf V}(t) \cdot {\bf V} \rangle,
\ee
where {\bf V} is the center of mass velocity of the vesicle motor in $d$-dimension. The center of mass velocity can be decomposed in the averaged directed velocity in the ${\bf \hat R}(t)$ direction and fluctutations as, ${\bf V} = {\bf \hat R}(t) \langle V_z \rangle + \delta $ giving 
\be
D_{e} = D_0 + \frac{1}{d}{\langle V_{z} \rangle}^2 \tau _R.
\ee

The first term on the right hand side is the diffusion coefficient in absence of any propulsion, while the second term is characterized by the decay of the orientational correlation function with an orientational relaxation time $\tau_R$.  The diffusion coefficient can equivalently determined from the mean square displacement (MSD)  $\Delta L^2(t) = \langle |{\bf r}_{CM}(t) - {\bf r}_{CM}(0)|^2\rangle$ as $D_e = lim_{t \rightarrow \infty} \Delta L^2(t)/dt$.  There are two relevant characteristic time scales in the system~\cite{golestanian:09}. First is the characteristic diffusion time of the product solvent particles around the vesicle given by $\tau_D= R_v/D$, where D is the  diffusion coefficient of solvent molecules and $R_v$ is the vesicle size. This time scale sets the relaxation time of the redistribution of the particles around the vesicle when it changes orientation.  Second is the orientational relaxation time $\tau_R$ which controls the changes in the orientation of the spherical vesicle, and is defined via the orientation autocorrelation function: $C_\theta(t) =  \langle {\bf \hat R}(t) \cdot {\bf \hat R}(0)\rangle \sim e^{-(t/\tau_R) }$.  The solvent transport coefficient can be calculated analytically for MPC dynamics~\cite{snigdha:12}, giving $\tau_D \approx 600$ for vesicle with $N_v = 110$.

\begin{figure}
\begin{center}
\includegraphics[scale=0.5]{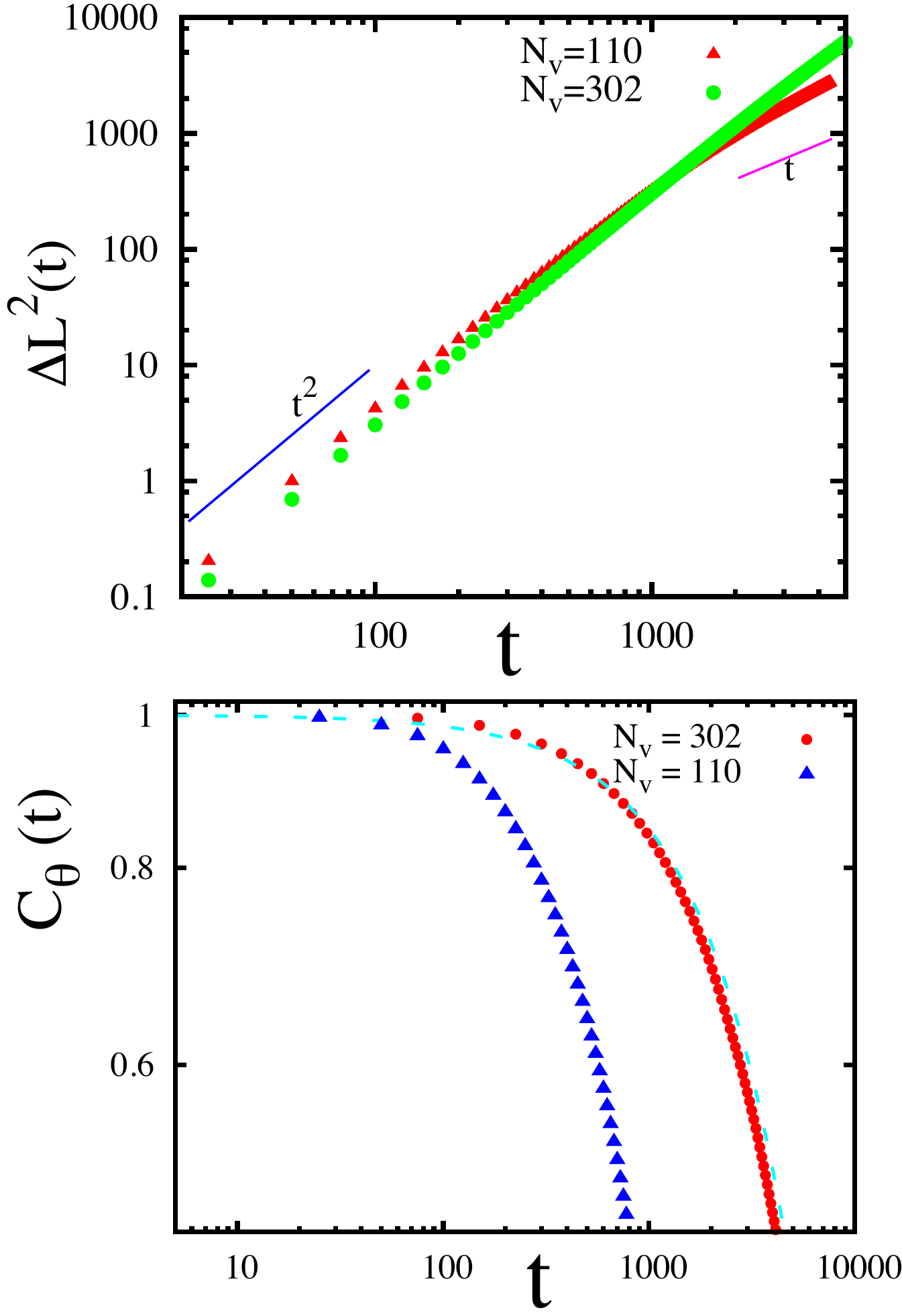}
\caption{\label{fig:msd} Top: Mean square displacement $\Delta L^2(t)$ for different sized vesicles having $C_f=25\%$ depicting the ballistic motion $\sim t^2$ at short times and diffusive motion $\sim t$ at long times. Bottom: Orientational autocorrelation function for two vesicles. The dashed line is an exponential fit to data.}
\end{center}
\end{figure}

For short times ($t \ll \tau_D$) MSD is shown to exhibit ballistic regime~\cite{golestanian:09} where $\Delta L ^2(t) \sim \langle V_z \rangle^2 t^2$, while for long times $t \gg \tau_R$ it exhibits a linear behaviour as $\Delta L ^2(t) \sim 6(D_0 + \langle V_z \rangle^2 \tau_R/d)t$. Fig.~\ref{fig:msd}(top) shows the  MSD for the vesicle of different sizes $N_v = 110$ and $302$. Fitting of the short and long time of this plot is comparable with the theoretical expression given above to estimate the  average vesicle propagation velocity $\langle V_z \rangle$ and reorientation time $\tau_R$ and enhanced diffusion coefficient.  Fig.~\ref{fig:msd}(bottom) shows the orientation autocorrelation function for different sized vesicle.  Table~\ref{I} shows the variation of  these quantities by varying the size of vesicle.
\begin{center}
\begin{table}[htbp]
\begin{tabular}[t]{c c  c c}
\hline
$N_v$ \ \ \ & \ \  \ $\langle V_z \rangle$ \ \ \ & \ \ \ $\tau_R$ \ \ \  &  \ \ \ $D_e$ \\
\hline
  29  \ \ \ &   \ \ \    0.021 \ \ \ & \ \ \ 175  \ \ \  & \ \ \ 0.020\\
  110  \ \ \ &   \ \ \    0.019 \ \ \ & \ \ \ 1100  \ \ \  & \ \ \ 0.132\\
  194 \ \ \ & \ \ \ 0.018 \ \ \ & \ \ \ 2500 \ \ \ & \ \ \  0.270 \\
  302 \ \ \ & \ \ \  0.017 \ \ \  & \ \ \  6000 \ \ \ & \ \ \ 0.578 \\
  434 \ \ \ & \ \ \  0.016 \ \ \ & \ \ \ 11500 \ \  \ & \ \ \  0.981 \\
  \hline
	\hline
\end{tabular}
\caption{The average directed velocity, orientational relaxation time and the enhanced diffusion coefficients, obtained from the mean square displacement and orientation autocorrelation function for  different sizes of spherical vesicles.}
\label{I}
\end{table}
\end{center}

It is instructive to consider the values of these quantities for motors with various size. For most of the vesicle sizes reported in table~\ref{I} we find the diffusion coefficient in absence of propulsion to be $D_0 \sim 10^{-3}$, whereas the contribution from self-propulsion in $D_e$ is much higher. Consequently, the dynamics for these vesicles are dominated by the directed motion. However, as decrease the vesicle size the directed motion contribution given by $\frac{1}{3}\langle V_z \rangle ^2 \tau_R$ keeps decreasing and, for very small vesicle motors with $N_v=29$, $D_e$ is much smaller. This is due to the fact that the translational Brownian motion contribution starts to dominate and directed motion is compromised.  One can therefore infer here that the orientational Brownian motion is not only stochastic element, rather it alters the dynamics of chemically powered motor significantly.

\subsection*{Prolate/Ellipsoidal vesicle}

Vesicles are highly adaptive structures having a rich diversity of shapes. Some important reasons that drives the shape transition are temperature, pressure, curvature energy~\cite{sackmann-91,seifert}etc. In order to get an insight on the propulsion efficiency of vesicle in various shapes, we induce the shape change either by changing the number density  of solvent inside the vesicle, which in turn changes the reduced volume $v$ and hence the shape of the vesicle~\cite{li-09} or by varying the bending elasticity and thus by curvature.  

\begin{figure}[h]
\begin{center}
\includegraphics[scale=0.4]{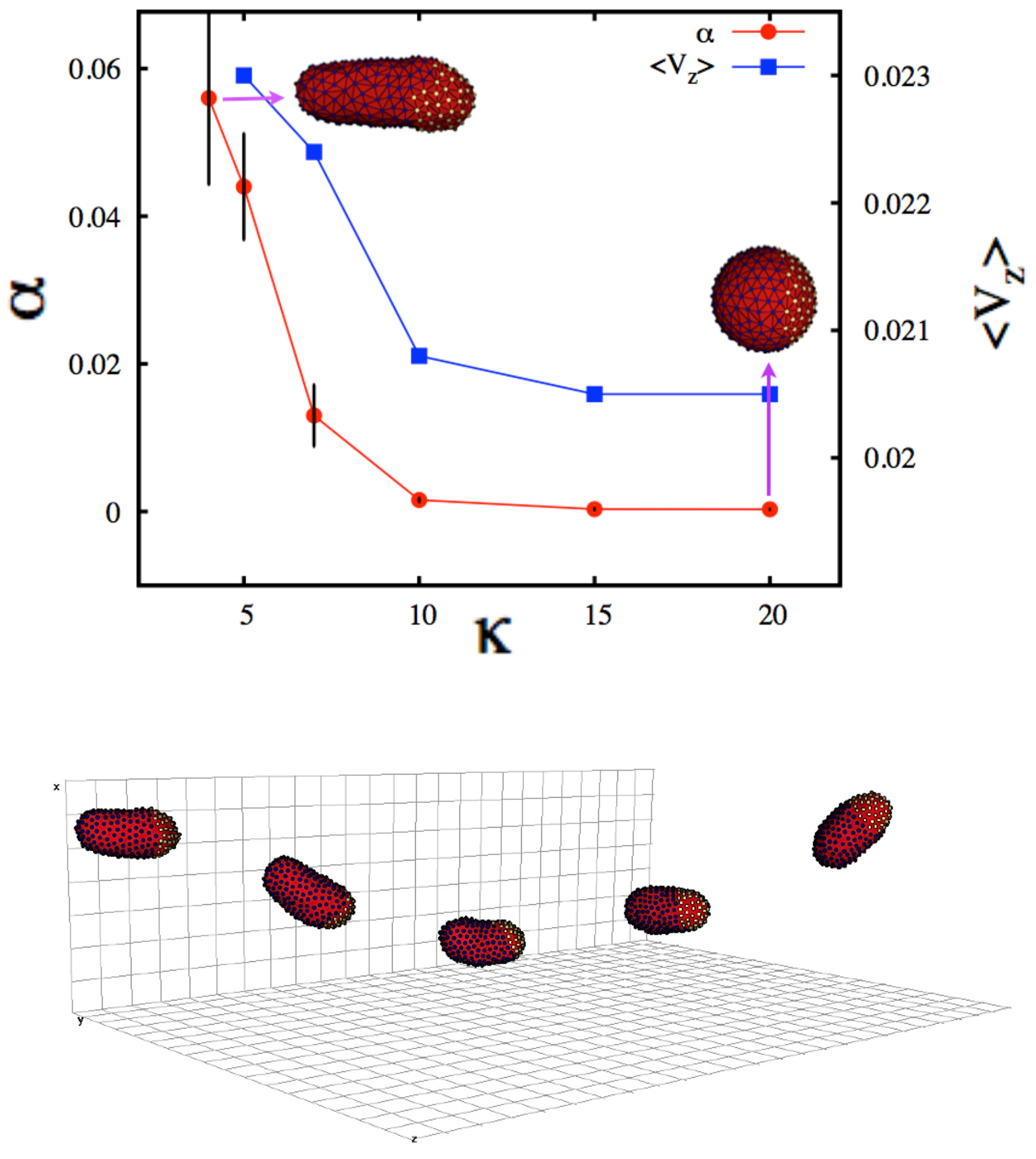}
\caption{\label{fig:aspher} Top:  Variation of asphericity $\alpha$ and the average directed velocity $\langle V_z \rangle$ as a function of bending rigidity $\kappa$ keeping $\rho_{in} = 1$. Bottom: The propulsion and reorientation of a prolate vesicle.}
\end{center}
\end{figure}

\begin{figure}[h]
\begin{center}
\includegraphics[scale=0.24]{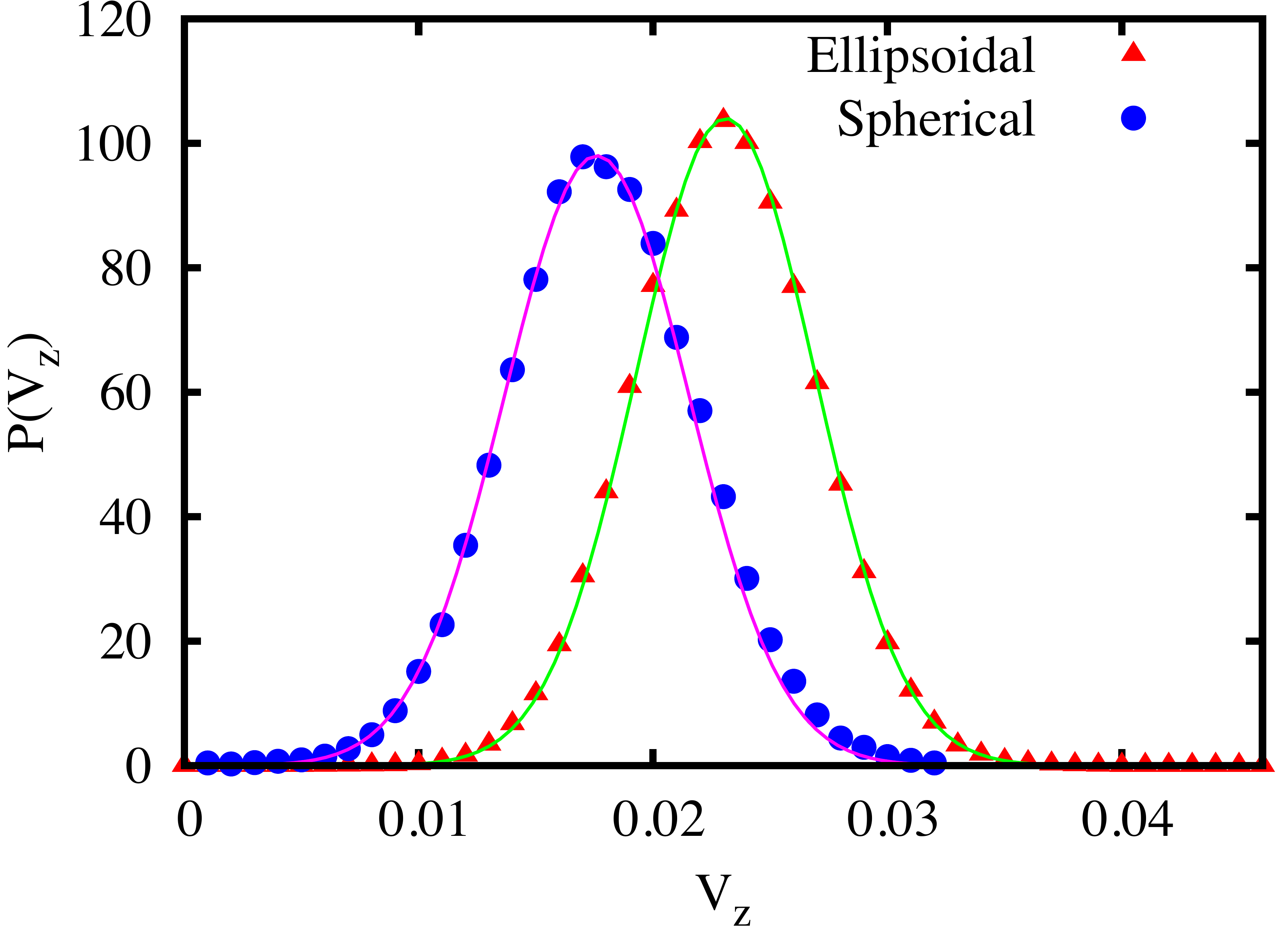}
\caption{\label{fig:ellipvz}Plot of $P(V_z)$, the probability density of $V_z$ for vesicle with $N_v=302$ with prolate and spherical shape. Green and pink solid lines are maxwell boltzmann fit to data.}
\end{center}
\end{figure}

To evaluate the morphological changes of the vesicle in a quantitative manner, we calculated "asphericity", $\alpha$ as an order parameter that reflects the deviation from the spherical shape~\cite{gompper-05}. Asphericity $\alpha$ is obtained from the  three eigenvalues ($\lambda_1,\lambda_2,\lambda_3$) of the moment-of-inertia tensor given as
\be
\alpha = \frac{(\lambda_1-\lambda_2)^2 + (\lambda_2-\lambda_3)^2 + (\lambda_3-\lambda_1)^2}{2R_g^4}
\ee
 where, $R_g^2 = \lambda_1 + \lambda_2 + \lambda_3$. Fig.~\ref{fig:aspher}(Top) shows the increase in asphericity on reducing the bending rigidity $\kappa$ of the vesicle. The observed shape transition in this case is from spherical vesicle to prolate/ellipsoidal vesicle. This shape change is accompanied with a strong change in dynamical behaviour of the vesicle. Fig.~\ref{fig:aspher}(Bottom) shows a motile prolate which reorients in due course of time due to thermal fluctuation. Fig.~\ref{fig:ellipvz} shows the comparison of the probability distribution of directed propulsive velocity for spherical and ellipsoidal vesicle. It is quite evident that the prolate/ellipsoidal vesicle has enhanced propulsion velocity,$\langle V_z \rangle = 0.023$ as compared to the spherical shape, $\langle V_z \rangle = 0.017$.  
It must be noted here the motion of the elliptical vesicle is such that the catalytic vertices are around the tip of the ellipse and movement is along the long axis. This orientation can be explained in terms of reduced bending which allows the solvent interactions to deform the vesicle thus reducing the drag.

The increased propulsion velocity in case of prolate/ellipsoidal vesicle can be attributed to the less drag experienced by the vesicle from its environment in this shape. The strength of the drag force depends not only on the viscosity at low
speeds, but also on the cross-sectional shape that is presented to the fluid by the object in its direction of motion. For a spherical vesicle of radius $R$ the drag coefficient is given by the Stoke's law as $f_t = 6 \pi \eta R$, where $\eta$ is the viscosity of the medium. For an ellipsoidal case the drag coefficient changes to $f_t = \frac{4 \pi \eta a}{ln(2a/b) - 1/2}$ for motion at low speed parallel to the long axis of the ellipsoid~\cite{boal-book}. Here $a$ and $b$ are semi-major and semi-minor axis respectively. The calculated drag coefficient for the vesicle having $N_v = 302$ in prolate and spherical shapes are found out to be $\approx 98$ and $125$ respectively. We believe that, it is this decrease in drag coefficient which results in enhancement of the directed velocity of the prolate vesicle.

Does the increased directed velocity of the ellipsoidal vesicle guarantee it to be a better candidate for microscale swimmer than a spherical vesicle? To answer this next we probed the mean square displacement and orientational angular correlation for the prolate shape and compared it with the sphere. 

\begin{figure}[h]
\begin{center}
\includegraphics[scale=0.37]{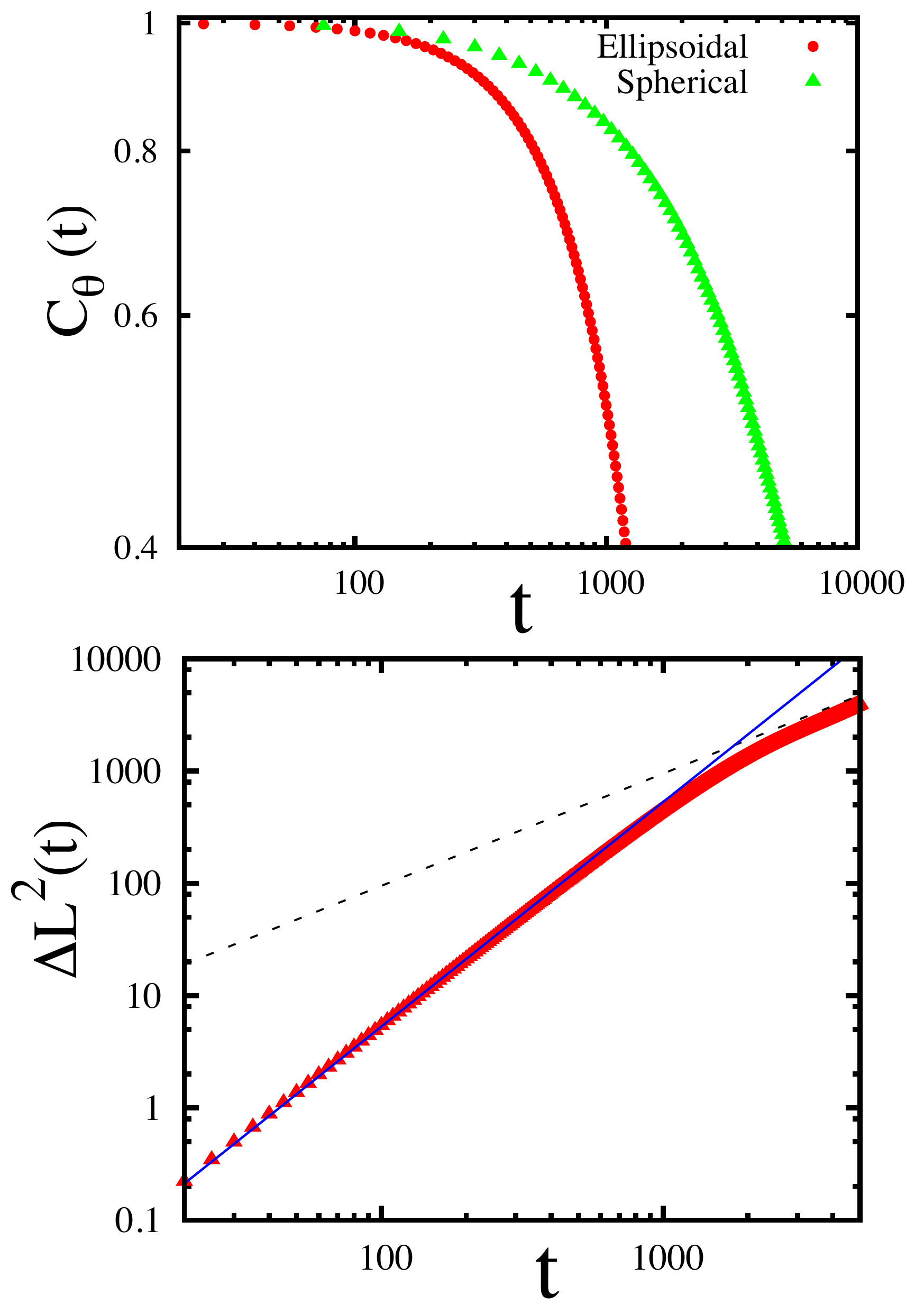}
\caption{\label{fig:ellip-msd}Top:  Orientational autocorrelation function as a function of time for  ellipsoidal and spherical vesicle having $N_v = 302$. Bottom: Mean square displacement $\Delta L^2(t)$ for ellipsoidal vesicles having $N_v = 302$ depicting the ballistic motion $\sim t^2$ at short times and diffusive motion $\sim t$ at long times. }
\end{center}
\end{figure}

Fig.~\ref{fig:ellip-msd} (Top) compares the orientation autocorrelation function $C_\theta(t)$ for the two different shapes discussed so far, which can be used to calculate the orientational relaxation time $\tau_R$. The faster decay of autocorrelation function in case of ellipsoidal vesicle clearly depicts the faster reorientation of the vesicle in prolate shape.  The relaxation time $\tau_R$  for the prolate shape is found out to be 1700, which is   much smaller than the spherical vesicle having same number of vertices. Table~\ref{I} shows the spherical vesicle of $N_v=302$ has $\tau_R=6000$. Further, the change of directed motion to random walk at the time scale of $\tau_R$ is also shown in the fig.~\ref{fig:ellip-msd}(Bottom). The reduction in the reorientation time for ellipsoidal case could be due increase in floppiness in the prolate vesicle and hence that of the catalytic tip.  
In case the prolate vesicle would have been a rigid object, then the concentration gradient produced by the vesicle that is responsible for the self propulsion would have been along the vector ${\bf \hat R}(t)$ and would have been symmetric normal to the vector. However, presence of floppiness breaks this symmetry and induces forces that lead to reorientation of the vesicle. 

\subsection*{Discocyte vesicle}

\begin{figure}[h]
\begin{center}
\includegraphics[width=1.0\columnwidth]{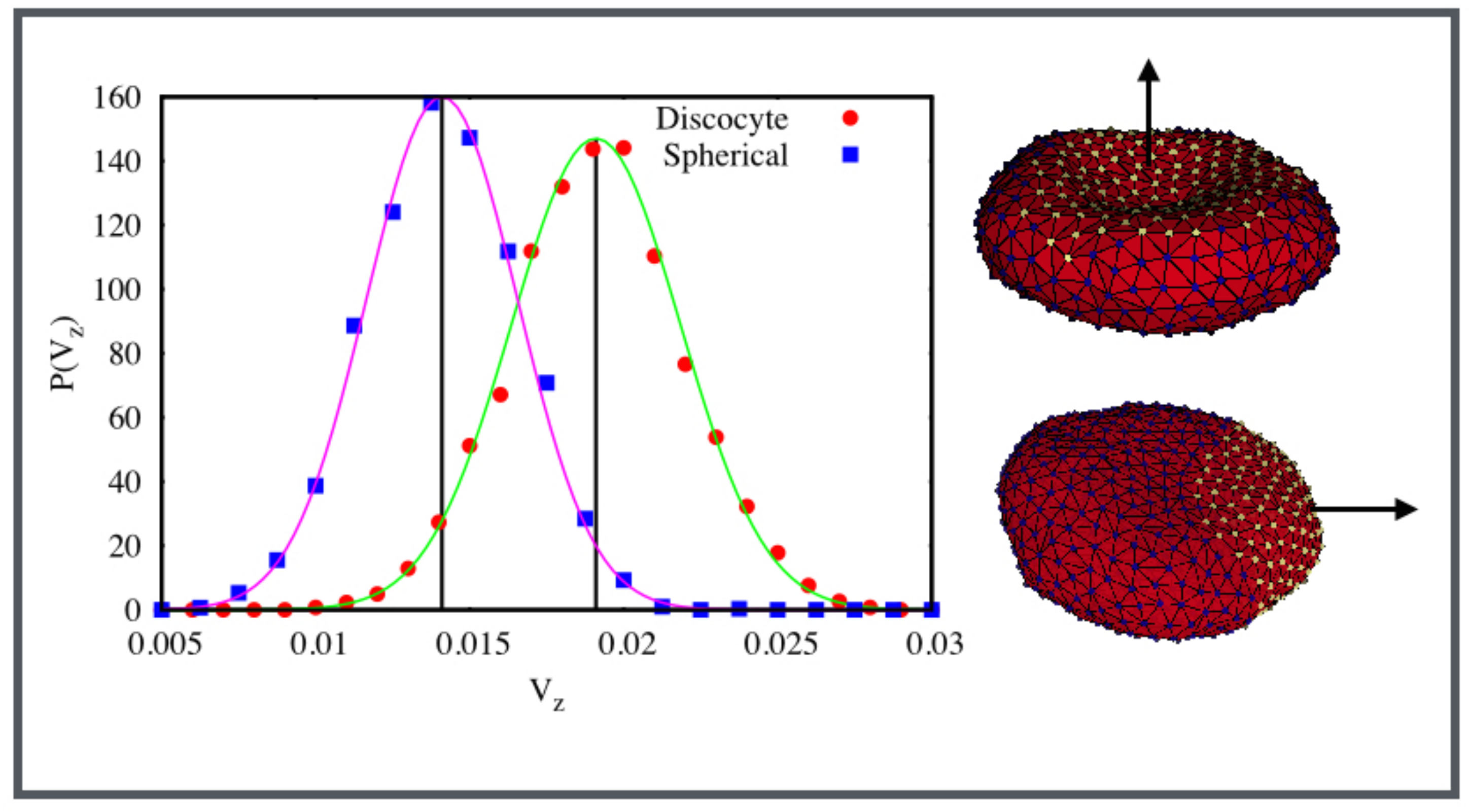}
\caption{\label{fig:discocyte} Left Panel: Comparision of $P(V_z)$, the probability density of $V_z$ for vesicle with $N_v=677$ in discocyte and spherical shape. Right Panel: Instantaneous configuration of moving discocytes in the face on and  edge-on mode. Arrow shows the direction of movement of the vesicle.}
\end{center}
\end{figure}

As stated before, changing the reduced volume $v$ leads to various shapes of the vesicle. Next we investigate the propulsion of vesicles having a discocyte shape, which was obtained by reducing the number density of solvent inside the vesicle to $\rho _{in} = 2$ with $\kappa = 5$  . Unlike the case of ellipsoids, where we have a unique mode of propulsion with the velocity direction along its long axis, here we obtain two different modes of propulsion, in one the discocyte moves with its face on, and in the other it moves with its edge at the front. Fig.~\ref{fig:discocyte}(right panel) shows both the propulsion configuration.
Fig.~\ref{fig:discocyte} (left panel) compare the directed velocity $V_z$ for the discocyte with a spherical vesicle having same vertex number $N_v = 677$. It is evident from the plot that the discocyte have higher propulsion with $\langle V_z \rangle= 0.019$ than its spherical counterpart with  $\langle V_z \rangle= 0.014$. 

\section {Vesicle Motor Efficiency}
\label{sec-eff}
Chemically powered vesicle motors converts chemical energy into mechanical work, driving the self-propelled directed motion, while working in an environment where viscous drag is important. 
The efficiency $\eta _s$ of chemically-powered motor, which measures the effectiveness of mechanochemical energy transduction, can be defined as~\cite{wang-02, wang:13}

\be
\eta_s = \frac{P_{mech}}{P_{chem}} = \frac{\zeta\langle V_z \rangle^2}{\Delta \mu R}
\ee
where, $P_{mech}$ is the mechanical power output of the motor and $P_{chem}$ is the total chemical power input to the motor. $P_{mech}$ and $P_{chem}$ are further defined as $P_{mech} = \zeta \langle V_z \rangle^2$ and $P_{chem} = \Delta \mu R$, where $\zeta$ is the drag coefficient of the vesicle motor, $\langle V_z \rangle$ is the directed propulsion velocity of the vesicle,  $R$ is the net chemical reaction rate and $\Delta \mu$ is the change in the chemical potential in the reaction. To determine the efficiency from above equation, the net chemical reaction rate $R$ was calculated by counting the number of $A \rightarrow B$ reactive events that take place at the catalytic vertices per unit time. In the MPC-MD simulations the change in the chemical potential is given by 
\be
\Delta \mu = \mu_B - \mu_A = -K_BT \ln \frac{n_B}{n_B^{eq}}\frac{n_A^{eq}}{n_A},
\ee

where $n_A^{eq}$ and $n_B^{eq}$ denote the equilibrium number densities of $A$ and $B$ species respectively, while $n_A$ and $n_B$ are the steady state densities~\cite{kapral-book}. The drag coefficient depends not only on the viscosity, but also on the cross-sectional shape that is presented to the fluid by the vesicle in its direction of motion. For spheres, $\zeta = 6 \pi \eta R$, for ellipsoids moving along its long axis $\zeta =  \frac{4 \pi \eta a}{ln(2a/b) - 1/2}$ and for disks moving edge on  $\zeta = \frac{32}{3} \eta p$, where $p$ is the radius of the disk. We compare the Stokes efficiency of the various sizes and shapes of the vesicle in Table~\ref{II}

\begin{center}
\begin{table}[htbp]
\begin{tabular}[t]{c c c c c}
\hline
$N_v$ \ \ \ & \ \ \ Shape \ \ \ & \ \ \  $\epsilon_B$ \ \ \ & \ \ \ $\langle V_z \rangle$ \ \ \ & \ \ \ $\eta_s$  \\
\hline
  110  \ \ \ &   \ \ \    Spherical \ \ \ & \ \ \ 0.01 \ \ \ & \ \ \ 0.019 \ \ \ & \ \ \  0.0027 \\
   302  \ \ \ &   \ \ \    Spherical \ \ \ & \ \ \ 0.01 \ \ \ & \ \ \ 0.017 \ \ \ & \ \ \  0.0020 \\
   302  \ \ \ &   \ \ \    Spherical \ \ \ & \ \ \ 0.1 \ \ \ & \ \ \ 0.008 \ \ \ & \ \ \  0.0004 \\
   302  \ \ \ &   \ \ \    Ellipsoidal \ \ \ & \ \ \ 0.01 \ \ \ & \ \ \ 0.023 \ \ \ & \ \ \  0.0030 \\
   677  \ \ \ &   \ \ \    Discocyte \ \ \ & \ \ \ 0.01 \ \ \ & \ \ \ 0.019 \ \ \ & \ \ \  0.0015 \\
   677  \ \ \ &   \ \ \   Spherical \ \ \ & \ \ \ 0.01 \ \ \ & \ \ \ 0.014 \ \ \ & \ \ \  0.0012 \\
  \hline
	\hline
\end{tabular}
\caption{Comparison of the average directed velocity and Stoke's efficiency for vesicle motor of various size and shapes. The LJ energy parameter $\epsilon_A = 1.0$.}
\label{II}
\end{table}
\end{center}

It is evident from the above calculation that the power associated with the mechanical work of the motor is larger for smaller vesicle of same shape, as the Stokes efficiency for spherical vesicle of $N_v=110$ is $0.27\%$ and that of vesicle with $N_v=302$ is $0.20\%$. On changing the shape of the vesicle its efficiency also changes. Table~\ref{II} shows that the ellipsoidal motor with $N_v=302$ is more efficient with $\eta_s=0.30\%$ than its spherical counterpart with $\eta_s=0.20\%$. Similarly for comparison of spherical and Discocyte shape shows enhancement of efficiency in the Discocyte shape. The maximum efficiency achieved in our simulations is about $0.30\%$, which is similar to the other chemically powered motors~\cite{yuguo-09,wang:13}, however the this efficiency is much less than that of most nano- and micromotors in biology~\cite{block-98}. 
\section{Conclusion} \label{sec:conc}
\label{sec-conc}

The particle-based mesoscopic model for chemically powered self-propelled vesicle dynamics discussed in
this article provides insight into the factors that control the motion. Our model couples the vesicle motion to the solvent by intermolecular forces, which ultimately manifest themselves in properties such as the surface tension of macroscopic scales. Here, we show that depending on the bending rigidity and solvent density around the vesicle, different shapes of vesicle can be obtained and all these shapes exhibit very different dynamical behavior.

Many potential applications of such vesicle motors will involve launching them to perform a given task, such as cargo transport, and further investigations of  interaction between them can provide the information needed to design motors that can cooperate with each other to perform such tasks.

{\em Acknowledgments}: The work of S.T. was supported by Department of Science and Technology, India. The computation work was carried out at the HPC facility in IISER Bhopal, India. 

\bibliographystyle{prsty}

\begin{thebibliography}{}

\end{thebibliography}


\begin{thebibliography}{10}

\bibitem{golestanian:05}
R. Golestanian, T.~B. Liverpool, and A. Ajdari, Phys. Rev. Lett. {\bf 94},
  220801  (2005).

\bibitem{nardi:99}
J. Nardi, R. Bruinsma, and E. Sackmann, Phys. Rev. Lett. {\bf 82},  5168
  (1999).

\bibitem{sumino:05}
Y. Sumino, N. Magome, T. Hamada, and K. Yoshikawa, Phys. Rev. Lett. {\bf 94},
  068301  (2005).

\bibitem{kay:07}
E.~R. Kay, D.~A. Leigh, and F. Zerbetto, Angew. Chem. Int. Ed. {\bf 46},  72
  (2007).

\bibitem{anderson:83}
J.~L. Anderson, Phys. Fluids {\bf 26},  2871  (1983).

\bibitem{vale:00}
R.~D. Vale and R.~A. Milligan, Science {\bf 288},  88  (2000).

\bibitem{yildiz:06}
A. Yildiz, Science {\bf 311},  792  (2006).

\bibitem{shin:04}
J.-S. Shin and N.~A. Pierce, Journal of the American Chemical Society {\bf
  126},  10834  (2004), pMID: 15339155.

\bibitem{shapere:87}
S.-P. at~Low Reynolds~Number, Phys. Rev. Lett. {\bf 58},  2051  (1987).

\bibitem{sen:04}
W.~F. Paxton {\it et~al.}, J. Am. Chem. Soc. {\bf 126},  13424  (2004).

\bibitem{wang:13}
W. Wang, T.-Y. Chiang, D. Velegol, and T.~E. Mallouk, J. Am. Chem. Soc. {\bf
  135},  10557  (2013).

\bibitem{golestanian-1:07}
J.~R. Howse {\it et~al.}, Phys. Rev. Lett. {\bf 99},  048102  (2007).

\bibitem{ozin:10}
L.~F. Valadares {\it et~al.}, Small {\bf 6},  565  (2010).

\bibitem{ruckner:07}
G. R\"uckner and R. Kapral, Phys. Rev. Lett. {\bf 98},  150603  (2007).

\bibitem{snigdha:12}
S. Thakur and R. Kapral, Phys. Rev. E {\bf 85},  026121  (2012).

\bibitem{wu:13}
Z. Wu {\it et~al.}, Angew. Chem. Int. Ed. {\bf 52},  7000  (2013).

\bibitem{wilson-12}
R.~J. M.~N. Daniela A.~Wilson and J.~C.~M. van Hest, Nature Chemistry {\bf 4},
  268  (2012).

\bibitem{howse-12}
J. Howse, Nature Chemistry {\bf 4},  247  (2012).

\bibitem{ohta:09}
T. Ohta and T. Ohkuma, Phys. Rev. Lett. {\bf 102},  154101  (2009).

\bibitem{sano-13}
H.~T. Miki Y.~Matsuo and M. Sano, EPL {\bf 102},  40012  (2013).

\bibitem{miura:09}
T. Miura {\it et~al.}, Langmuir {\bf 26},  1610  (2010).

\bibitem{giardini:03}
P.~A. Giardini, D.~A. Fletcher, and J.~A. Theriot, PNAS {\bf 100},  6493
  (2003).

\bibitem{ohta-12}
A. Menzel and T. Ohta, EPL {\bf 99},  58001  (2012).

\bibitem{kapral_adv_chem_phys_2008}
R. Kapral, Adv. Chem. Phys. {\bf 140},  89  (2008).

\bibitem{gompper:09}
G. Gompper, T. Ihle, D.~M. Kroll, and R.~G. Winkler, Adv. Polym. Sci. {\bf
  221},  1  (2009).

\bibitem{malevanets-99}
A. Malevanets and R. Kapral, J. Chem. Phys. {\bf 110},  8605  (1999).

\bibitem{malevanets-00}
A. Malevanets and R. Kapral, J. Chem. Phys. {\bf 112},  72609  (2000).

\bibitem{ho-89}
J.-S. Ho and A. Baumg\"artner, Phys. Rev. Lett. {\bf 63},  1324  (1989).

\bibitem{nelson-book}
G. Gompper and D.~M. Kroll, {\em Mechanics of membrane and surfaces, ed. D. R.
  Nelson, T. Piran and S. Weinberg} (World Scientific, Singapore, 2004).

\bibitem{noguchi-05}
H. Noguchi and G. Gompper, PNAS {\bf 102},  14159  (2005).

\bibitem{gompper-05}
H. Noguchi and G. Gompper, PRE {\bf 72},  011901  (2005).

\bibitem{helfrich-73}
W. Helfrich, Z. Naturforsch C. {\bf 28},  693  (1973).

\bibitem{weber-85}
F.~H. Stillinger and T.~A. Weber, PRB {\bf 31},  5262  (1985).

\bibitem{snigdha-14}
D. Sarkar, S. Thakur, Y.-G. Tao, and R. Kapral, Soft Matter {\bf 10},  9577
  (2014).

\bibitem{kroll-01}
T. Ihle and D.~M. Kroll, Phys. Rev. E {\bf 63},  020201  (2001).

\bibitem{kroll-03}
T. Ihle and D.~M. Kroll, Phys. Rev. E {\bf 67},  066705  (2003).

\bibitem{golestanian:07}
R. Golestanian, T.~B. Liverpool, and A. Ajdari, New J. Phys. {\bf 9},  126
  (2007).

\bibitem{snigdha:10}
S. Thakur and R. Kapral, J. Chem. Phys. {\bf 133},  204505  (2010).

\bibitem{yuguo-08}
Y.-G. Tao and R. Kapral, J. Chem. Phys. {\bf 128},  164518  (2008).

\bibitem{golestanian:09}
R. Golestanian, Phys. Rev. Lett. {\bf 102},  188305  (2009).

\bibitem{sackmann-91}
J. Kas and E. Sackmann, Biophys. Journal {\bf 60},  825  (1991).

\bibitem{seifert}
U. Seifert, Adv. Phys. {\bf 46},  13  (1997).

\bibitem{li-09}
X. Li, I.~V. Pivkin, H. Liang, and G.~E. Karniadakis, Macromolecules {\bf 42},
  3195  (2009).

\bibitem{boal-book}
D. Boal, {\em Mechanics of the Cell} (Cambridge University Press, Cambridge,
  2012).

\bibitem{wang-02}
H. Wang and G. Oster, Europhys. Lett. {\bf 57},  134  (2002).

\bibitem{kapral-book}
R. Kapral,  in {\em Engineering of Chemical Complexity} (World Scientific
  Publishing Co. Pte. Ltd., Singapore, 2013), Chap.~5, pp.\ 101--124.

\bibitem{yuguo-09}
Y.-G. Tao and R. Kapral, J. Chem. Phys. {\bf 131},  024113  (2009).

\bibitem{block-98}
S.~M. Block, Cell {\bf 93},  5  (1998).

\end{thebibliography}

\end{document}